\documentclass[conference]{IEEEtran}
\usepackage{ifpdf}
\usepackage[nospace]{cite}
\usepackage{graphicx}
\usepackage{color}
\usepackage{bm}
\usepackage[cmex10]{amsmath}
\usepackage{amssymb}
\usepackage{array}
\usepackage{syntonly}
\usepackage{amsthm}
\usepackage{amsfonts}
\usepackage{epsfig}
\usepackage{latexsym}

\usepackage{fixltx2e}
\usepackage{url}
\theoremstyle{definition} 
\theoremstyle{definition} 
\theoremstyle{definition} 
\begin{document}
\title{{\LARGE Transmitter Optimization in MISO Broadcast Channel with Common and Secret Messages} }
\author{{\large Sanjay Vishwakarma and A. Chockalingam} \\
Email: sanjay@ece.iisc.ernet.in, achockal@ece.iisc.ernet.in \\
{\normalsize Department of ECE, Indian Institute of Science,
Bangalore 560012, India}
}
\maketitle
\begin{abstract}
In this paper, we consider transmitter optimization in multiple-input single-output (MISO) 
broadcast channel with common and secret messages.
The secret message is intended for $K$ users and it is transmitted with perfect secrecy with respect to 
$J$ eavesdroppers which are also assumed 
to be legitimate users in the network.
The common message is transmitted at a fixed rate $R_{0}$ and 
it is intended for all $K$ users and $J$ eavesdroppers.
The source operates under a total power constraint. It also injects artificial noise
to improve the secrecy rate.
We obtain the optimum covariance matrices associated with
the common message, secret message, and artificial noise,
which maximize the achievable secrecy rate and simultaneously meet the fixed rate $R_{0}$ for the common message.
\end{abstract}
{\em keywords:}
{\em {\footnotesize
Physical layer security, MISO, common and secret messages, secrecy rate, artificial noise, multiple eavesdroppers. 
}} 
\IEEEpeerreviewmaketitle

\section{Introduction}
\label{sec1}
The concept of achieving perfect secrecy using physical layer techniques
was first introduced in \cite{ic0} on a degraded wiretap channel.
Later, this work was extended to more general broadcast channel and Gaussian channel
in \cite{ic1} and \cite{ic2}, respectively.
Achieving secrecy using physical layer techniques as opposed to cryptographic techniques
does not rely on the computational limitation of the eavesdroppers.
Wireless networks can be easily eavesdropped due to the broadcast
nature of the information transmission. With the growing applications
on wireless networks, there is a growing demand for achieving secrecy on these networks.
Secrecy in multi-antenna point-to-point wireless links has been studied by several
authors, e.g., \cite{ic3, ic4, ic11, ic5, ic6}.
In \cite{ic1}, simultaneous transmission of a private message to receiver 1 
at rate $R_{1}$ and a common message to receivers 1 and 2 at rate $R_{0}$ 
for two discrete memoryless channels with common input was considered. 
Recently, the work in \cite{ic1} has been extended to multiple-input multiple-output 
(MIMO) broadcast channel 
with confidential and common messages in \cite{ic7,ic8,ic9}. 
Motivated by the above works, in this paper, we consider transmitter optimization
in multiple-input single-output (MISO) broadcast channel 
with common and secret messages.
The secret message is intended for $K$ users and it is transmitted with perfect secrecy with respect to
$J$ eavesdroppers which are also assumed 
to be legitimate users in the network.
The common message is transmitted at a fixed rate $R_{0}$ and 
it is intended for all $K$ users and $J$ eavesdroppers.
The source operates under a total power constraint. It 
also injects artificial noise to improve the secrecy rate.
Under these settings, we obtain the optimum covariance matrices associated with
the common message, secret message, and artificial noise,
which maximize the achievable secrecy rate and simultaneously meet the 
fixed rate $R_{0}$ for the common message.
We note that the secrecy rate maximization in MISO channel without common message and in the presence of
multiple eavesdroppers has been considered in \cite{ic20} where the secret message is intended
only for a single user (i.e., $K = 1$). 

$\bf{Notations:}$ $\boldsymbol{A} \in 
\mathbb{C}^{N_{1} \times N_{2}}$ implies that $\boldsymbol{A}$ is a complex 
matrix of dimension $N_{1} \times N_{2}$. 
$\boldsymbol{A} \succeq \boldsymbol{0}$ implies
that $\boldsymbol{A}$ is a positive semidefinite matrix.
Complex conjugate 
transpose operation is denoted by $[.]^{\ast}$.
$\mathbb{E}[.]$ denotes the expectation operator, and
$\parallel.\parallel$ denotes the 2-norm operator.
\section{System Model}
\label{sec2}
\begin{figure}
\center
\includegraphics[totalheight=6.5cm,width=6.5cm]{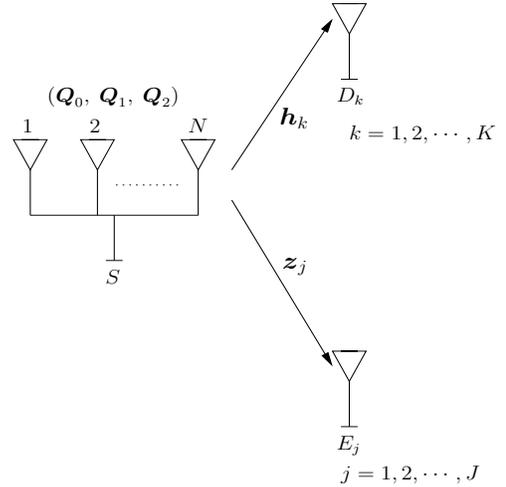}
\caption{System model for MISO broadcast channel with common and secret messages.} 
\label{fig1}
\end{figure}

\vspace{0.0mm}
Consider a MISO broadcast channel as shown in Fig. \ref{fig1}
which consists 
of a source $S$ having $N$ transmit antennas, $K$ users $ \{ D_{1}, D_{2},\cdots,D_{K}\}$ each 
having single antenna, and $J$ eavesdroppers $ \{ E_{1}, E_{2},\cdots,E_{J}\}$ each 
having single antenna. The complex channel gain from $S$ to $D_{k}$ is 
denoted by $\boldsymbol{h}_{k} \in \mathbb{C}^{1 \times N}$, $1\leq k \leq K$.
Likewise, the complex channel gain from $S$ to $E_{j}$ is denoted by 
$\boldsymbol{z}_{j} \in \mathbb{C}^{1 \times N}$, $1\leq j \leq J$.
We assume that eavesdroppers are non-colluding \cite{ic6}.

Let $P_{T}$ denote the total transmit power budget in the system, i.e., the source
$S$ operates under total power constraint $P_{T}$.
The communication between the source and the users and eavesdroppers
happens in $n$ channel uses. The source $S$ transmits two 
independent messages $W_{0}$ and $W_{1}$, which are equiprobable over 
$\{1,2,\cdots,2^{nR_{0}} \}$ and $\{1,2,\cdots,2^{nR_{1}} \}$, respectively. 
$W_{0}$ is the common message to be conveyed to all $D_{k}$s and $E_{j}$s at 
information rate $R_{0}$. $W_{1}$ is 
the secret message which has to be conveyed to all $D_{k}$s at some rate 
$R_{1}$ with perfect secrecy with respect to all $E_{j}$s \cite{ic6}. 
For each $W_{0}$ drawn equiprobably from the set 
$\{1,2,\cdots,2^{nR_{0}} \}$, the source maps $W_0$ to an i.i.d. 
$(\sim \mathcal{CN}(\boldsymbol{0},\boldsymbol{Q}_{0}))$ codeword $\{ \boldsymbol{X}^{0}_{i} \}^{n}_{i = 1}$ 
of length $n$, where each $\boldsymbol{X}^{0}_{i} \in \mathbb{C}^{N\times1}$ and 
$\boldsymbol{Q}_{0} = \mathbb{E}[ \boldsymbol{X}^{0}_{i} \boldsymbol{X}^{0\ast}_{i} ]$. 
Similarly, for each $W_{1}$ drawn equiprobably from the set 
$\{1,2,\cdots,2^{nR_{1}} \}$, the source, using a stochastic encoder,
maps $W_{1}$ to an i.i.d. $(\sim \mathcal{CN}(\boldsymbol{0},\boldsymbol{Q}_{1}))$ codeword 
$\{ \boldsymbol{X}^{1}_{i} \}^{n}_{i = 1}$
of length $n$, where each $\boldsymbol{X}^{1}_{i} \in \mathbb{C}^{N\times1}$ and 
$\boldsymbol{Q}_{1} = \mathbb{E}[ \boldsymbol{X}^{1}_{i} \boldsymbol{X}^{1\ast}_{i} ]$.
The source also injects i.i.d. $(\sim \mathcal{CN}(\boldsymbol{0},\boldsymbol{Q}_{2}))$ artificial noise sequence 
$\{ \boldsymbol{X}^{2}_{i} \}^{n}_{i = 1}$ of length $n$,
where each $\boldsymbol{X}^{2}_{i} \in \mathbb{C}^{N\times1}$ and 
$\boldsymbol{Q}_{2} = \mathbb{E}[ \boldsymbol{X}^{2}_{i} \boldsymbol{X}^{2\ast}_{i} ]$.
In the $i$th channel use, $1 \leq i \leq n$, the source transmits the sum of the symbols which is
$\boldsymbol{X}^{0}_{i} + \boldsymbol{X}^{1}_{i} + \boldsymbol{X}^{2}_{i}$.
Since the source is power limited, this implies that 
%
\begin{eqnarray}
trace(\boldsymbol{Q}_{0}) + trace(\boldsymbol{Q}_{1}) + trace(\boldsymbol{Q}_{2})\ \leq \ P_{T}. \label{eqn1}
\end{eqnarray}
%
In the following, 
we will use $\boldsymbol{X}^{0}$, $\boldsymbol{X}^{1}$ and $\boldsymbol{X}^{2}$ to denote the symbols in the codewords 
$\{ \boldsymbol{X}^{0}_{i} \}^{n}_{i = 1}$ and 
$\{ \boldsymbol{X}^{1}_{i} \}^{n}_{i = 1}$, and the artificial noise sequence
$\{ \boldsymbol{X}^{2}_{i} \}^{n}_{i = 1}$,
respectively. We also assume that all the channel gains are known and remain static over the codeword 
transmit duration.
Let $y_{D_{k}}$ and $y_{E_{j}}$ denote the received signals at $D_{k}$ and $E_{j}$, respectively.
We have
%
\begin{eqnarray}
y_{D_{k}} \ = \ \boldsymbol{h}_{k} \boldsymbol{X}^{0} + \boldsymbol{h}_{k} \boldsymbol{X}^{1} + \boldsymbol{h}_{k} \boldsymbol{X}^{2} + \eta_{D_{k}}, \nonumber \\ \forall k = 1,2,\cdots,K, \label{eqn2} \\
y_{E_{j}} \ = \ \boldsymbol{z}_{j} \boldsymbol{X}^{0} + \boldsymbol{z}_{j} \boldsymbol{X}^{1} + \boldsymbol{z}_{j} \boldsymbol{X}^{2} + \eta_{E_{j}}, \nonumber \\ \forall j = 1,2,\cdots,J, \label{eqn3}
\end{eqnarray}
%
where the $\eta$s are the noise components, assumed to be i.i.d. $\mathcal{CN}(0,N_0)$.
\section{Transmitter optimization in MISO broadcast channel}
\label{sec3}
Since the symbol $\boldsymbol{X}^{0}$ is transmitted at 
information rate $R_{0}$ irrespective of $\boldsymbol{X}^{1}$, treating $\boldsymbol{X}^{1}$ as noise in (\ref{eqn2}), 
$D_{k}$s will be able to decode $\boldsymbol{X}^{0}$ if $ \ \forall k = 1,2,\cdots,K,$
%
\begin{eqnarray}
I\big(\boldsymbol{X}^{0}; y^{}_{D_{k}}\big) \ = \ \log_{2}\Big(1 +  \frac{\boldsymbol{h}_{k}\boldsymbol{Q}_{0}\boldsymbol{h}^{\ast}_{k}}{N_{0} + \boldsymbol{h}_{k}\boldsymbol{Q}_{1}\boldsymbol{h}^{\ast}_{k} + \boldsymbol{h}_{k}\boldsymbol{Q}_{2}\boldsymbol{h}^{\ast}_{k}} \Big) \nonumber \\ \geq R_{0}. \label{eqn18}
\end{eqnarray}
%
Similarly, treating $\boldsymbol{X}^{1}$ as noise in (\ref{eqn3}), 
$E_{j}$s will be able to decode $\boldsymbol{X}^{0}$ if $ \ \forall j = 1,2,\cdots,J,$
%
\begin{eqnarray}
I\big(\boldsymbol{X}^{0}; y^{}_{E_{j}}\big)  \ =  \ \log_{2}\Big(1 +  \frac{\boldsymbol{z}_{j}\boldsymbol{Q}_{0}\boldsymbol{z}^{\ast}_{j}}{N_{0} + \boldsymbol{z}_{j}\boldsymbol{Q}_{1}\boldsymbol{z}^{\ast}_{j} + \boldsymbol{z}_{j}\boldsymbol{Q}_{2}\boldsymbol{z}^{\ast}_{j}} \Big) \nonumber \\ \geq R_{0}. \label{eqn19}
\end{eqnarray}
%
Using (\ref{eqn2}) and with the knowledge of the symbol $\boldsymbol{X}^{0}$, the information 
rate for $\boldsymbol{X}^{1}$ at $D_{k}$ is
%
\begin{eqnarray}
I\big(\boldsymbol{X}^{1}; y_{D_{k}} \mid \boldsymbol{X}^{0} \big) = \log_{2} \Big( 1 + \frac{\boldsymbol{h}_{k}\boldsymbol{Q}_{1} \boldsymbol{h}^{\ast}_{k}}{N_{0} + \boldsymbol{h}_{k}\boldsymbol{Q}_{2} \boldsymbol{h}^{\ast}_{k}} \Big). \label{eqn20}
\end{eqnarray}
%
Similarly, using (\ref{eqn3}) and with the knowledge of $\boldsymbol{X}^0$, 
the information rate for $\boldsymbol{X}^{1}$ at $E_{j}$ is
%
\begin{eqnarray}
I\big(\boldsymbol{X}^{1}; y_{E_{j}} \mid \boldsymbol{X}^{0} \big) = \log_{2} \Big( 1 + \frac{\boldsymbol{z}_{j}\boldsymbol{Q}_{1} \boldsymbol{z}^{\ast}_{j}}{N_{0} + \boldsymbol{z}_{j}\boldsymbol{Q}_{2} \boldsymbol{z}^{\ast}_{j}} \Big). \label{eqn21}
\end{eqnarray}
%
%
\subsection{Transmitter optimization - without artificial noise}
\label{subsec3A}
In this subsection, we consider transmitter optimization in MISO broadcast channel 
when no artificial noise is injected by the source.
Subject to the constraints in (\ref{eqn1}), (\ref{eqn18}) and 
(\ref{eqn19}), the achievable secrecy rate for $\boldsymbol{X}^{1}$ is obtained 
by solving the following optimization problem \cite{ic6}:
%
\begin{eqnarray}
R_{1} \ = \ \max_{\boldsymbol{Q}_{0}, \ \boldsymbol{Q}_{1}} \ \min_{k=1,2,\cdots,K \atop{j= 1,2,\cdots,J}} \nonumber \\ \Big \{ I\big( \boldsymbol{X}^{1}; y_{D_{k}} \mid \boldsymbol{X}^{0} \big) - I \big(\boldsymbol{X}^{1}; y_{E_{j}} \mid \boldsymbol{X}^{0} \big) \Big \}  \label{eqn22} \\
= \ \max_{\boldsymbol{Q}_{0}, \ \boldsymbol{Q}_{1}} \min_{k=1,2,\cdots,K \atop{j = 1,2,\cdots,J}} \log_{2} \Big( \frac{1 + \frac{\boldsymbol{h}_{k}\boldsymbol{Q}_{1} \boldsymbol{h}^{\ast}_{k}}{N_{0}}}{1 + \frac{\boldsymbol{z}_{j}\boldsymbol{Q}_{1} \boldsymbol{z}^{\ast}_{j}}{N_{0}}} \Big) \label{eqn37} \\
= \ \log_{2} \max_{\boldsymbol{Q}_{0}, \ \boldsymbol{Q}_{1}} \min_{k=1,2,\cdots,K \atop{j=1,2,\cdots,J}} \Big( \frac{N_{0} + \boldsymbol{h}_{k}\boldsymbol{Q}_{1} \boldsymbol{h}^{\ast}_{k}}{N_{0} + \boldsymbol{z}_{j}\boldsymbol{Q}_{1} \boldsymbol{z}^{\ast}_{j}} \Big) \label{eqn38}
\end{eqnarray}
s.t.
\begin{eqnarray}
\log_{2}\Big(1 +  \frac{\boldsymbol{h}_{k}\boldsymbol{Q}_{0}\boldsymbol{h}^{\ast}_{k}}{N_{0} + \boldsymbol{h}_{k}\boldsymbol{Q}_{1}\boldsymbol{h}^{\ast}_{k}} \Big) \ \geq \ R_{0}, \ \forall k = 1,2,\cdots,K,  \label{eqn23}\\
\log_{2}\Big(1 +  \frac{\boldsymbol{z}_{j}\boldsymbol{Q}_{0}\boldsymbol{z}^{\ast}_{j}}{N_{0} + \boldsymbol{z}_{j}\boldsymbol{Q}_{1}\boldsymbol{z}^{\ast}_{j}} \Big) \ \geq \ R_{0}, \ \forall j = 1,2,\cdots,J,  \label{eqn24}\\
\boldsymbol{Q}_{0} \succeq \boldsymbol{0}, \quad \boldsymbol{Q}_{1} \succeq \boldsymbol{0}, \quad trace(\boldsymbol{Q}_{0}) + trace(\boldsymbol{Q}_{1}) \ \leq \ P_{T}. \label{eqn25}
\end{eqnarray}
%
The constraints (\ref{eqn23}) and (\ref{eqn24}) are obtained from 
(\ref{eqn18}) and (\ref{eqn19}), respectively. The objective function 
in (\ref{eqn22}) is obtained from (\ref{eqn20}) and (\ref{eqn21}). 
We rewrite the optimization problem in (\ref{eqn38}) in the following equivalent form:
%
\begin{eqnarray}
\max_{\boldsymbol{Q}_{0}, \ \boldsymbol{Q}_{1}} \ \min_{k = 1,2,\cdots,K \atop{j = 1,2,\cdots,J}} \ \Big( \frac{N_{0} + \boldsymbol{h}_{k}\boldsymbol{Q}_{1} \boldsymbol{h}^{\ast}_{k}}{N_{0} + \boldsymbol{z}_{j}\boldsymbol{Q}_{1} \boldsymbol{z}^{\ast}_{j}} \Big) \label{eqn39} 
\end{eqnarray}
s.t.
\begin{eqnarray}
\Big(1 +  \frac{\boldsymbol{h}_{k}\boldsymbol{Q}_{0}\boldsymbol{h}^{\ast}_{k}}{N_{0} + \boldsymbol{h}_{k}\boldsymbol{Q}_{1}\boldsymbol{h}^{\ast}_{k}} \Big) \ \geq \ 2^{R_{0}}, \ \forall k = 1,2,\cdots,K,  \nonumber \\
\Big(1 +  \frac{\boldsymbol{z}_{j}\boldsymbol{Q}_{0}\boldsymbol{z}^{\ast}_{j}}{N_{0} + \boldsymbol{z}_{j}\boldsymbol{Q}_{1}\boldsymbol{z}^{\ast}_{j}} \Big) \ \geq \ 2^{R_{0}}, \ \forall j = 1,2,\cdots,J,  \nonumber \\
\boldsymbol{Q}_{0} \succeq \boldsymbol{0}, \quad \boldsymbol{Q}_{1} \succeq \boldsymbol{0}, \quad trace(\boldsymbol{Q}_{0}) + trace(\boldsymbol{Q}_{1}) \ \leq \ P_{T}. \label{eqn40}
\end{eqnarray}
%
Further, we rewrite the innermost minimization in (\ref{eqn39}), namely,
%
\begin{eqnarray}
\min_{k = 1,2,\cdots,K \atop{j = 1,2,\cdots,J}} \ \left( \frac{N_{0} + \boldsymbol{h}_{k}\boldsymbol{Q}_{1} \boldsymbol{h}^{\ast}_{k}}{N_{0} + \boldsymbol{z}_{j}\boldsymbol{Q}_{1} \boldsymbol{z}^{\ast}_{j}} \right), \label{eqn41}  
\end{eqnarray}
%
in the following equivalent maximization form:
%
\begin{eqnarray}
 \max_{t} \ t \label{eqn42} \\
\text{s.t.} \quad  t\big(N_{0} + \boldsymbol{z}_{j}\boldsymbol{Q}_{1} \boldsymbol{z}^{\ast}_{j} \big) - \big( N_{0} + \boldsymbol{h}_{k}\boldsymbol{Q}_{1} \boldsymbol{h}^{\ast}_{k} \big) \ \leq \ 0, \nonumber \\ \forall k = 1,2,\cdots,K, \quad \forall j = 1,2,\cdots,J. \label{eqn43}
\end{eqnarray}
%
Substituting the above maximization form in (\ref{eqn39}), we get the following single maximization form:
%
\begin{eqnarray}
\max_{\boldsymbol{Q}_{0}, \ \boldsymbol{Q}_{1}, \ t} \ \ t \label{eqn44} \\
\text{s.t.} \quad 
\forall k = 1,2,\cdots,K, \quad \forall j = 1,2,\cdots,J, \nonumber \\
t\big(N_{0} + \boldsymbol{z}_{j}\boldsymbol{Q}_{1} \boldsymbol{z}^{\ast}_{j} \big) - \big( N_{0} + \boldsymbol{h}_{k}\boldsymbol{Q}_{1} \boldsymbol{h}^{\ast}_{k} \big) \ \leq \ 0, \nonumber \\
\big(2^{R_{0}} - 1 \big) \big( N_{0} + \boldsymbol{h}_{k}\boldsymbol{Q}_{1}\boldsymbol{h}^{\ast}_{k} \big)  - \big( \boldsymbol{h}_{k}\boldsymbol{Q}_{0}\boldsymbol{h}^{\ast}_{k} \big) \ \leq \ 0, \nonumber \\
\big(2^{R_{0}} - 1 \big) \big( N_{0} + \boldsymbol{z}_{j}\boldsymbol{Q}_{1}\boldsymbol{z}^{\ast}_{j} \big)  - \big( \boldsymbol{z}_{j}\boldsymbol{Q}_{0}\boldsymbol{z}^{\ast}_{j} \big) \ \leq \ 0, \nonumber \\
\boldsymbol{Q}_{0} \succeq \boldsymbol{0}, \quad \boldsymbol{Q}_{1} \succeq \boldsymbol{0}, \quad trace(\boldsymbol{Q}_{0}) + trace(\boldsymbol{Q}_{1}) \ \leq \ P_{T}.
\label{eqn45} 
\end{eqnarray}
%
For a given $t$, the above problem is formulated as the following semidefinite
feasibility problem \cite{ic10}: 
%
\begin{eqnarray}
\text{find} \quad \boldsymbol{Q}_{0}, \ \boldsymbol{Q}_{1}, \label{eqn46}
\end{eqnarray}
%
subject to the constraints in (\ref{eqn45}). 
The maximum value of $t$, denoted by $t_{max}$, can be obtained using 
bisection method as follows. Let $t^{}_{max}$ lie in the interval 
$[t_{ll}, t_{ul}]$. The value of $t_{ll}$ can be taken as 1 (corresponding 
to the minimum secrecy rate of 0) and $t_{ul}$ can be taken as 
$(1 + \min_{k = 1,2,\cdots,K} \frac{P_{T} {\parallel \boldsymbol{h}_{k} \parallel}^{2}}{N_{0}})$, which corresponds to the 
minimum information capacity among $D_{k}$s when the entire power $P_{T}$ is 
allotted to the source $S$. 
Check the feasibility of 
(\ref{eqn45}) at $t^{} = (t^{}_{ll} + t^{}_{ul})/2$. If feasible, then 
$t^{}_{ll} = t^{}$, else $ \ t^{}_{ul} = t^{}$. Repeat this until 
$t^{}_{ul} - t^{}_{ll} \leq \zeta$, where $\zeta$ is a small positive 
number. Using $t^{}_{max}$ in (\ref{eqn38}), the secrecy rate is given by
%
\begin{eqnarray}
R_{1} = \log_2 t^{}_{max}. \label{eqn47}
\end{eqnarray}
%
{\em Remark:} \
We note that the maximum common message information rate, $R^{max}_{0}$, can be obtained as follows:
%
\begin{eqnarray}
R^{max}_{0} = \max_{\boldsymbol{Q}_{0}} \ \min_{k = 1,2,\cdots,K \atop{j = 1,2,\cdots,j}} \  \big \{ I\big(\boldsymbol{X}^{0}; \ y^{}_{D_{k}}\big), \ I \big(\boldsymbol{X}^{0}; \ y^{}_{E_{j}} \big) \big \} \label{eqn66} \\
\text{s.t.} \quad \boldsymbol{Q}_{0} \succeq \boldsymbol{0}, \quad trace(\boldsymbol{Q}_{0}) \ \leq \ P_{T}, \label{eqn67}
\end{eqnarray}
%
where $I\big(\boldsymbol{X}^{0}; \ y^{}_{D_{k}} \big)$ and $I \big(\boldsymbol{X}^{0}; \ y^{}_{E_{j}} \big)$ 
in (\ref{eqn66}) are obtained from (\ref{eqn18}) and (\ref{eqn19}), respectively, with 
$\boldsymbol{Q}_{1} = \boldsymbol{Q}_{2} = \boldsymbol{0}$. 
The above optimization problem can be easily solved using the method as proposed above to solve (\ref{eqn38}).
Also, using the K.K.T conditions, it can be shown that $R^{max}_{0}$ attains its maximum
value when $trace(\boldsymbol{Q}_{0}) = P_{T}$, i.e., when all the available power is used.
This implies that for $R_{1} > 0$, $R_{0} < R^{max}_{0}$.
\subsection{Rank-1 approximation of $\boldsymbol{Q}_{1}$ and $\boldsymbol{Q}_{0}$ - without artificial noise}
\label{subsec3B}
The optimal solutions $\boldsymbol{Q}_{0}$ and $\boldsymbol{Q}_{1}$ obtained from (\ref{eqn44}) may or may not have rank 1.
This can be easily seen from the K.K.T conditions of the optimization problem (\ref{eqn44}).
For practical application, a rank-1 approximation of $\boldsymbol{Q}_{0}$ and $\boldsymbol{Q}_{1}$ can be done as follows.
Let $\boldsymbol{\phi}^{0} \in \mathbb{C}^{N \times 1}$ and $\boldsymbol{\phi}^{1} \in \mathbb{C}^{N \times 1}$
be the unit norm eigen directions of $\boldsymbol{Q}_{0}$ and $\boldsymbol{Q}_{1}$ 
corresponding to the largest eigen values, respectively.
We take $P_{0}\boldsymbol{\phi}^{0}\boldsymbol{\phi}^{0\ast}$ and $P_{1}\boldsymbol{\phi}^{1}\boldsymbol{\phi}^{1\ast}$
as the rank-1 approximation of $\boldsymbol{Q}_{0}$ and $\boldsymbol{Q}_{1}$, respectively,
where $P_{0} \geq 0$, $P_{1} \geq 0$ and $P_{0} + P_{1} \ \leq P_{T}$. We substitute 
$\boldsymbol{Q}_{0} = P_{0}\boldsymbol{\phi}^{0}\boldsymbol{\phi}^{0\ast}$
and  $\boldsymbol{Q}_{1} = P_{1}\boldsymbol{\phi}^{1}\boldsymbol{\phi}^{1\ast}$ in the
optimization problem (\ref{eqn44}), which results in the following optimization problem:
%
{\small
\begin{eqnarray}
\max_{P_{0}, \ P_{1}, \ t} \ \ t \label{eqn48} \\
\text{s.t.} \quad \quad 
\forall k = 1,2,\cdots,K, \quad \forall j = 1,2,\cdots,J, \nonumber \\
t\big(N_{0} + P_{1}\boldsymbol{z}_{j}\boldsymbol{\phi}^{1}\boldsymbol{\phi}^{1\ast} \boldsymbol{z}^{\ast}_{j} \big) - \big( N_{0} + P_{1}\boldsymbol{h}_{k}\boldsymbol{\phi}^{1}\boldsymbol{\phi}^{1\ast} \boldsymbol{h}^{\ast}_{k} \big) \leq 0, \nonumber \\
\big(2^{R_{0}} - 1 \big) \big( N_{0} + P_{1}\boldsymbol{h}_{k}\boldsymbol{\phi}^{1}\boldsymbol{\phi}^{1\ast}\boldsymbol{h}^{\ast}_{k} \big)  - \big(P_{0} \boldsymbol{h}_{k}\boldsymbol{\phi}^{0}\boldsymbol{\phi}^{0\ast}\boldsymbol{h}^{\ast}_{k} \big) \leq 0, \nonumber \\
\big(2^{R_{0}} - 1 \big) \big( N_{0} + P_{1}\boldsymbol{z}_{j}\boldsymbol{\phi}^{1}\boldsymbol{\phi}^{1\ast}\boldsymbol{z}^{\ast}_{j} \big)  - \big(P_{0} \boldsymbol{z}_{j}P_{0}\boldsymbol{\phi}^{0}\boldsymbol{\phi}^{0\ast}\boldsymbol{z}^{\ast}_{j} \big) \leq 0, \nonumber \\
P_{0}\geq 0, \quad P_{1}\geq 0, \quad P_{0} + P_{1} \ \leq P_{T}.
\label{eqn49} 
\end{eqnarray}
}

\vspace{-4mm}
\hspace{-5mm}
For a given $t$, the above problem is formulated as the following linear
feasibility problem: 
%
\begin{eqnarray}
\text{find} \quad P_{0}, \ P_{1}, \label{eqn50}
\end{eqnarray}
%
subject to the constraints in (\ref{eqn49}). The maximum value of $t$ can be obtained using the bisection method 
and the corresponding secrecy rate can be 
obtained using (\ref{eqn47}).
\subsection{Transmitter optimization - with artificial noise}
\label{subsec3C}
In this subsection, we consider transmitter optimization in MISO broadcast channel 
when artificial noise is injected by the source.
Subject to the constraints in (\ref{eqn1}), (\ref{eqn18}) and 
(\ref{eqn19}), the achievable secrecy rate for $\boldsymbol{X}^{1}$ is obtained 
by solving the following optimization problem: 

\vspace{-4mm}
{\small
\begin{eqnarray}
R_{1} \ = \ \max_{\boldsymbol{Q}_{0}, \ \boldsymbol{Q}_{1}, \ \boldsymbol{Q}_{2}} \ \min_{k=1,2,\cdots,K \atop{j=1,2,\cdots,J}} \nonumber \\ \Big \{ I\big( \boldsymbol{X}^{1}; y_{D_{k}} \mid \boldsymbol{X}^{0} \big) - I \big( \boldsymbol{X}^{1}; y_{E_{j}} \mid \boldsymbol{X}^{0} \big) \Big \} \label{eqn51} \\
= \max_{\boldsymbol{Q}_{0}, \ \boldsymbol{Q}_{1}, \ \boldsymbol{Q}_{2}} \ \min_{k = 1,2,\cdots,K \atop{j = 1,2,\cdots,J}} \ \log_{2} \bigg( \frac{1 + \frac{\boldsymbol{h}_{k}\boldsymbol{Q}_{1} \boldsymbol{h}^{\ast}_{k}}{N_{0} + \boldsymbol{h}_{k}\boldsymbol{Q}_{2} \boldsymbol{h}^{\ast}_{k}}}{1 + \frac{\boldsymbol{z}_{j}\boldsymbol{Q}_{1} \boldsymbol{z}^{\ast}_{j}}{N_{0} + \boldsymbol{z}_{j}\boldsymbol{Q}_{2} \boldsymbol{z}^{\ast}_{j}}} \bigg) \label{eqn52} \\
= \ \log_{2} \ \max_{\boldsymbol{Q}_{0}, \ \boldsymbol{Q}_{1}, \ \boldsymbol{Q}_{2}} \ \min_{k = 1,2,\cdots,K \atop{j = 1,2,\cdots,J}} \nonumber \\ \Big( \frac{N_{0} + \boldsymbol{h}_{k}\boldsymbol{Q}_{2} \boldsymbol{h}^{\ast}_{k} + \boldsymbol{h}_{k}\boldsymbol{Q}_{1} \boldsymbol{h}^{\ast}_{k}}{N_{0} + \boldsymbol{h}_{k}\boldsymbol{Q}_{2} \boldsymbol{h}^{\ast}_{k}} \Big)  \Big( \frac{N_{0} + \boldsymbol{z}_{j}\boldsymbol{Q}_{2} \boldsymbol{z}^{\ast}_{j}}{N_{0} + \boldsymbol{z}_{j}\boldsymbol{Q}_{2} \boldsymbol{z}^{\ast}_{j} + \boldsymbol{z}_{j}\boldsymbol{Q}_{1} \boldsymbol{z}^{\ast}_{j}} \Big) \label{eqn53} 
\end{eqnarray}
}
s.t.
\begin{eqnarray}
\forall k = 1,2,\cdots,K, \quad \forall j = 1,2,\cdots,J, \nonumber \\
\log_{2}\Big(1 +  \frac{\boldsymbol{h}_{k}\boldsymbol{Q}_{0}\boldsymbol{h}^{\ast}_{k}}{N_{0} + \boldsymbol{h}_{k}\boldsymbol{Q}_{2}\boldsymbol{h}^{\ast}_{k} +\boldsymbol{h}_{k}\boldsymbol{Q}_{1}\boldsymbol{h}^{\ast}_{k}} \Big) \ \geq \ R_{0},   \label{eqn54}\\
\log_{2}\Big(1 +  \frac{\boldsymbol{z}_{j}\boldsymbol{Q}_{0}\boldsymbol{z}^{\ast}_{j}}{N_{0} + \boldsymbol{z}_{j}\boldsymbol{Q}_{2}\boldsymbol{z}^{\ast}_{j} + \boldsymbol{z}_{j}\boldsymbol{Q}_{1}\boldsymbol{z}^{\ast}_{j}} \Big) \ \geq \ R_{0}, \label{eqn55}\\
\boldsymbol{Q}_{0} \succeq \boldsymbol{0}, \quad \boldsymbol{Q}_{1} \succeq \boldsymbol{0}, \quad \boldsymbol{Q}_{2} \succeq \boldsymbol{0}, \nonumber \\ trace(\boldsymbol{Q}_{0}) + trace(\boldsymbol{Q}_{1}) + trace(\boldsymbol{Q}_{2}) \ \leq \ P_{T}, \label{eqn56}
\end{eqnarray}
%
where the constraints (\ref{eqn54}) and (\ref{eqn55}) are obtained from 
(\ref{eqn18}) and (\ref{eqn19}), respectively, and the objective function 
in (\ref{eqn51}) is obtained from (\ref{eqn20}) and (\ref{eqn21}). 
We rewrite the optimization problem in (\ref{eqn53}) in the following equivalent form:
%
\begin{eqnarray}
\max_{\boldsymbol{Q}_{0}, \ \boldsymbol{Q}_{1}, \ \boldsymbol{Q}_{2}} \ \min_{k = 1,2,\cdots,K \atop{j = 1,2,\cdots,J}} \Big( \frac{N_{0} + \boldsymbol{h}_{k}\boldsymbol{Q}_{2} \boldsymbol{h}^{\ast}_{k} + \boldsymbol{h}_{k}\boldsymbol{Q}_{1} \boldsymbol{h}^{\ast}_{k}}{N_{0} + \boldsymbol{h}_{k}\boldsymbol{Q}_{2} \boldsymbol{h}^{\ast}_{k}} \Big) \nonumber \\ \Big( \frac{N_{0} + \boldsymbol{z}_{j}\boldsymbol{Q}_{2} \boldsymbol{z}^{\ast}_{j}}{N_{0} + \boldsymbol{z}_{j}\boldsymbol{Q}_{2} \boldsymbol{z}^{\ast}_{j} + \boldsymbol{z}_{j}\boldsymbol{Q}_{1} \boldsymbol{z}^{\ast}_{j}} \Big) \label{eqn57} \\
\text{s.t.} \quad \forall k = 1,2,\cdots,K, \quad \forall j = 1,2,\cdots,J, \nonumber \\
\Big(1 +  \frac{\boldsymbol{h}_{k}\boldsymbol{Q}_{0}\boldsymbol{h}^{\ast}_{k}}{N_{0} + \boldsymbol{h}_{k}\boldsymbol{Q}_{2}\boldsymbol{h}^{\ast}_{k} + \boldsymbol{h}_{k}\boldsymbol{Q}_{1}\boldsymbol{h}^{\ast}_{k}} \Big) \ \geq \ 2^{R_{0}},  \nonumber \\
\Big(1 +  \frac{\boldsymbol{z}_{j}\boldsymbol{Q}_{0}\boldsymbol{z}^{\ast}_{j}}{N_{0} + \boldsymbol{z}_{j}\boldsymbol{Q}_{2}\boldsymbol{z}^{\ast}_{j} + \boldsymbol{z}_{j}\boldsymbol{Q}_{1}\boldsymbol{z}^{\ast}_{j}} \Big) \ \geq \ 2^{R_{0}},   \nonumber \\
\boldsymbol{Q}_{0} \succeq \boldsymbol{0}, \quad \boldsymbol{Q}_{1} \succeq \boldsymbol{0}, \quad \boldsymbol{Q}_{2} \succeq \boldsymbol{0}, \nonumber \\ trace(\boldsymbol{Q}_{0}) + trace(\boldsymbol{Q}_{1}) + trace(\boldsymbol{Q}_{2}) \ \leq \ P_{T}. \label{eqn58}
\end{eqnarray}
%
Further, we rewrite the innermost minimization in (\ref{eqn57}), namely,
%
\begin{eqnarray}
\min_{k = 1,2,\cdots,K \atop{j = 1,2,\cdots,J}} \ \Big( \frac{N_{0} + \boldsymbol{h}_{k}\boldsymbol{Q}_{2} \boldsymbol{h}^{\ast}_{k} + \boldsymbol{h}_{k}\boldsymbol{Q}_{1} \boldsymbol{h}^{\ast}_{k}}{N_{0} + \boldsymbol{h}_{k}\boldsymbol{Q}_{2} \boldsymbol{h}^{\ast}_{k}} \Big) \nonumber \\ \Big( \frac{N_{0} + \boldsymbol{z}_{j}\boldsymbol{Q}_{2} \boldsymbol{z}^{\ast}_{j}}{N_{0} + \boldsymbol{z}_{j}\boldsymbol{Q}_{2} \boldsymbol{z}^{\ast}_{j} + \boldsymbol{z}_{j}\boldsymbol{Q}_{1} \boldsymbol{z}^{\ast}_{j}} \Big), \label{eqn59}  
\end{eqnarray}
%
in the following equivalent maximization form:
%
\begin{eqnarray}
 \max_{u, \ v} \ \ uv \label{eqn60} \\
\text{s.t.} \quad \quad \forall k = 1,2,\cdots,K, \quad \forall j = 1,2,\cdots,J, \nonumber \\
u \big( N_{0} + \boldsymbol{h}_{k}\boldsymbol{Q}_{2} \boldsymbol{h}^{\ast}_{k} \big) \nonumber \\ - \ \big( N_{0} + \boldsymbol{h}_{k}\boldsymbol{Q}_{2} \boldsymbol{h}^{\ast}_{k} + \boldsymbol{h}_{k}\boldsymbol{Q}_{1} \boldsymbol{h}^{\ast}_{k} \big) \ \leq \ 0, \nonumber \\
v \big ( N_{0} + \boldsymbol{z}_{j}\boldsymbol{Q}_{2} \boldsymbol{z}^{\ast}_{j} + \boldsymbol{z}_{j}\boldsymbol{Q}_{1} \boldsymbol{z}^{\ast}_{j} \big) \nonumber \\ - \ \big ( N_{0} + \boldsymbol{z}_{j}\boldsymbol{Q}_{2} \boldsymbol{z}^{\ast}_{j}\big) \ \leq \ 0.
\label{eqn61}
\end{eqnarray}
%
Substituting the above maximization form in (\ref{eqn57}), we get the following single maximization form:
%
\begin{eqnarray}
\max_{\boldsymbol{Q}_{0}, \ \boldsymbol{Q}_{1}, \ \boldsymbol{Q}_{2}, \ u, \ v} \ \ uv \label{eqn62} 
\end{eqnarray}
\begin{eqnarray}
\text{s.t.} \quad \quad 
\forall k = 1,2,\cdots,K, \quad \forall j = 1,2,\cdots,J, \nonumber \\
\big(2^{R_{0}} -1 \big) \big( N_{0} + \boldsymbol{h}_{k}\boldsymbol{Q}_{2}\boldsymbol{h}^{\ast}_{k} + \boldsymbol{h}_{k}\boldsymbol{Q}_{1}\boldsymbol{h}^{\ast}_{k} \big) \nonumber \\ - \ \big( \boldsymbol{h}_{k}\boldsymbol{Q}_{0}\boldsymbol{h}^{\ast}_{k} \big) \ \leq \ 0, \nonumber \\
\big(2^{R_{0}} -1 \big) \big( N_{0} + \boldsymbol{z}_{j}\boldsymbol{Q}_{2}\boldsymbol{z}^{\ast}_{j} + \boldsymbol{z}_{j}\boldsymbol{Q}_{1}\boldsymbol{z}^{\ast}_{j} \big) \nonumber \\ - \ \big( \boldsymbol{z}_{j}\boldsymbol{Q}_{0}\boldsymbol{z}^{\ast}_{j} \big) \ \leq \ 0, \nonumber \\
u \big( N_{0} + \boldsymbol{h}_{k}\boldsymbol{Q}_{2} \boldsymbol{h}^{\ast}_{k} \big) \nonumber \\ - \ \big( N_{0} + \boldsymbol{h}_{k}\boldsymbol{Q}_{2} \boldsymbol{h}^{\ast}_{k} + \boldsymbol{h}_{k}\boldsymbol{Q}_{1} \boldsymbol{h}^{\ast}_{k} \big) \ \leq \ 0, \nonumber \\
v \big ( N_{0} + \boldsymbol{z}_{j}\boldsymbol{Q}_{2} \boldsymbol{z}^{\ast}_{j} + \boldsymbol{z}_{j}\boldsymbol{Q}_{1} \boldsymbol{z}^{\ast}_{j} \big) \nonumber \\ - \ \big ( N_{0} + \boldsymbol{z}_{j}\boldsymbol{Q}_{2} \boldsymbol{z}^{\ast}_{j}\big) \ \leq \ 0, \nonumber \\
\boldsymbol{Q}_{0} \succeq \boldsymbol{0}, \quad \boldsymbol{Q}_{1} \succeq \boldsymbol{0}, \quad \boldsymbol{Q}_{2} \succeq \boldsymbol{0}, \nonumber \\ trace(\boldsymbol{Q}_{0}) + trace(\boldsymbol{Q}_{1}) + trace(\boldsymbol{Q}_{2}) \ \leq \ P_{T}.
\label{eqn63}
\end{eqnarray}
%
From the constraints in (\ref{eqn63}), it is obvious that the upper bound for $u$ can be taken as 
{\small $\big(1 + \min_{k = 1,2,\cdots,K} \frac{P_{T} {\parallel \boldsymbol{h}_{k} \parallel}^{2}}{N_{0}} \big)$} and we denote it by $u_{max}$.
Similarly, the upper bound for $v$ can be taken as $1$ and we denote it by $v_{max}$.
We denote the optimum value of the optimization problem (\ref{eqn62}) 
by $u_{opt}v_{opt}$. For positive secrecy rate, $u_{max} \geq u_{opt} > 1, \ v_{max} \geq v_{opt} > 0$ and $u_{opt}v_{opt} > 1$.
We obtain $u_{opt}v_{opt}$ 
sequentially by increasing $u$ from $1$ towards $u_{max}$ in discrete steps of size $\vartriangle_{u} \ = (u_{max} - 1)/M $, 
where $M$ is a large positive integer, 
and finding the maximum $v$ such that the constraints in (\ref{eqn63}) are feasible and the product $uv$ is maximum. 
The algorithm to obtain $u_{opt}v_{opt}$ as follows.  \\ 
\_\_\_\_\_\_\_\_\_\_\_\_\_\_\_\_\_\_\_\_\_\_\_\_\_\_\_\_\_\_\_\_\_\_\_\_\_\_\_\_\_\_\_\_\_\\
$1.$ \textbf{for} ($i \ = \ 1\ : \ 1\ : \ M$)\\
$2.$ \textbf{begin} \\
$3.$ $u_{i} \ = \ 1 + (i \ * \ \vartriangle_{u})$ \\ 
$4.$ $v_{i} \quad = \quad \mathop{\max}\limits_{\boldsymbol{Q}_{0}, \ \boldsymbol{Q}_{1}, \ \boldsymbol{Q}_{2}, \ v, \atop{u = u_{i}}}  \  v \ \text{ s.t. to all constraints in (\ref{eqn63}).}  \nonumber \label{eqn64}$ \\ 
$5.$ $\textbf{if}\ (i = 1) \ \textbf{then} \ u_{opt} = u_{i}, \ v_{opt} = v_{i}$ \\
$6.$ $\textbf{elseif} \ (u_{opt}v_{opt} \ \leq \ u_{i}  v_{i}) \ \textbf{then} \ u_{opt} = u_{i}, \ v_{opt} = v_{i}$ \\
$7.$ $\textbf{else} \ u_{opt} = u_{opt}, \ v_{opt} = v_{opt}$ \\
$8.$ $\textbf{endif} $  \\ 
$9.$ $\textbf{end for loop}$ 
\hspace{-5mm} \\
\_\_\_\_\_\_\_\_\_\_\_\_\_\_\_\_\_\_\_\_\_\_\_\_\_\_\_\_\_\_\_\_\_\_\_\_\_\_\_\_\_\_\_\_\_

\vspace{2mm}
The constrained maximization problem in the for loop can be solved 
using the bisection method by checking the feasibility of the constraints in (\ref{eqn63}) at 
$u = u_{i}$ and $v$ in the interval $[0, v_{max}]$. 
Having obtained $u_{opt}v_{opt}$, the secrecy rate is given by
$R^{}_{1} = \log_2 u_{opt}v_{opt}. \label{eqn65}$

We can take the rank-1 approximation of $\boldsymbol{Q}_{1}$ and $\boldsymbol{Q}_{0}$ as discussed in subsection \ref{subsec3B},
i.e., by substituting $\boldsymbol{Q}_{0} = P_{0}\boldsymbol{\phi}^{0}\boldsymbol{\phi}^{0\ast}$
and  $\boldsymbol{Q}_{1} = P_{1}\boldsymbol{\phi}^{1}\boldsymbol{\phi}^{1\ast}$ in the
optimization problem (\ref{eqn62}) and solving for $P_{0}, \ P_{1}, \ \boldsymbol{Q}_{2}, \ u$ and $v$.
\section{Results and Discussions}
\label{sec5}
We present the numerical results and discussions in this section. We 
obtained the secrecy rate results through simulations for $N = 2$,
$K = 1$ and $J = 1,2,3$ eavesdroppers. The following complex channel gains are 
taken in the simulations:
$\boldsymbol{h}     = [2.0824 - 1.7215i,  \ 0.0788 - 0.0583i]$, 
$\boldsymbol{z}_{1} = [-0.3989 - 0.0923i, \  -0.6770 + 0.3371i]$,  
$\boldsymbol{z}_{2} = [0.0910 - 0.8258i,  \ 0.6642 - 0.3257i]$, 
$\boldsymbol{z}_{3} = [-0.2784 - 1.3995i, \ -1.4867 + 0.9877i]$.

\begin{figure}[htb]
\begin{minipage}[b]{.48\linewidth}
\centering
\centerline{\epsfig{figure=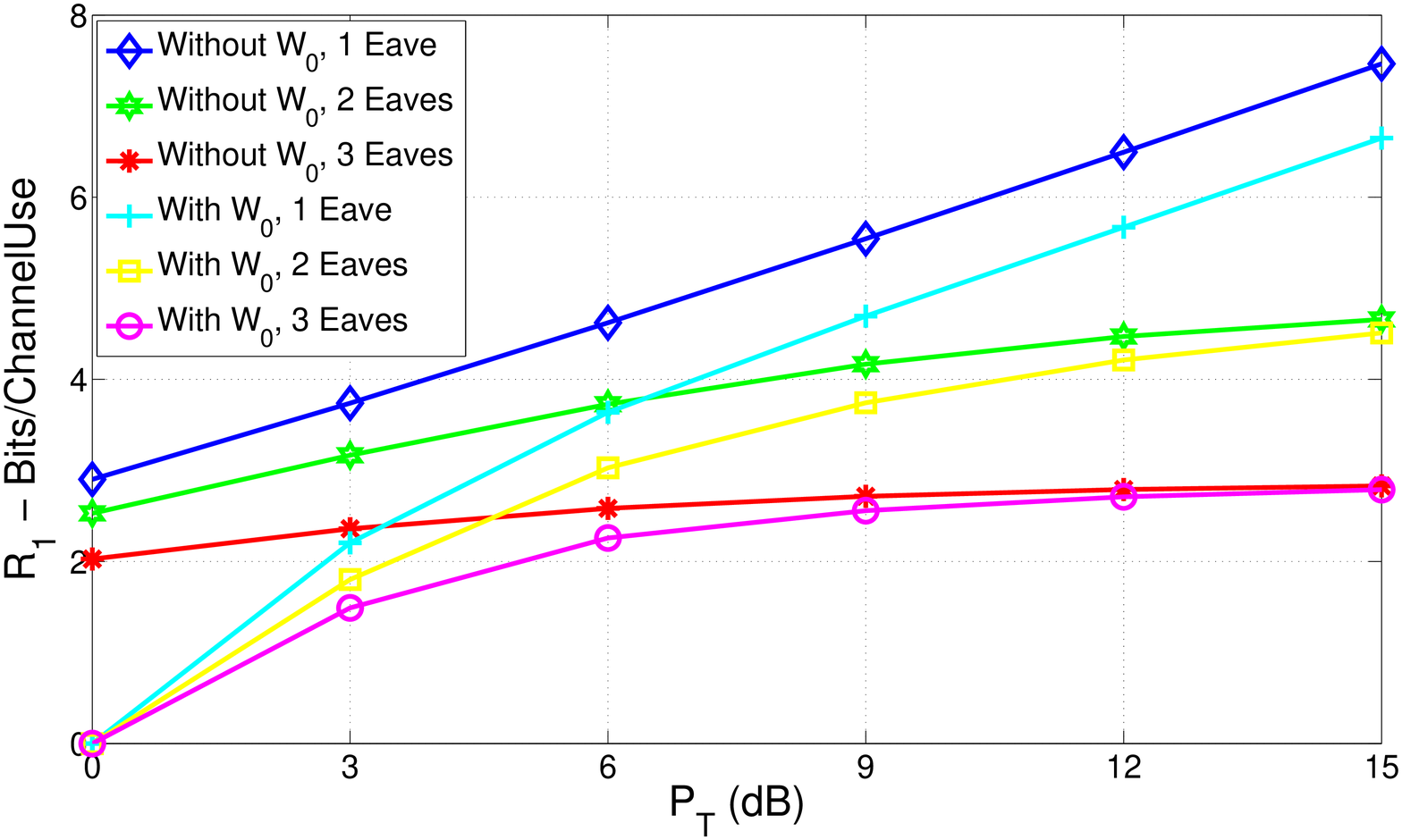,width=5.0cm,height=7.5cm}}
\centerline{\small{(a)}}\medskip
\end{minipage}
\hfill
\begin{minipage}[b]{0.48\linewidth}
\centering
\centerline{\epsfig{figure=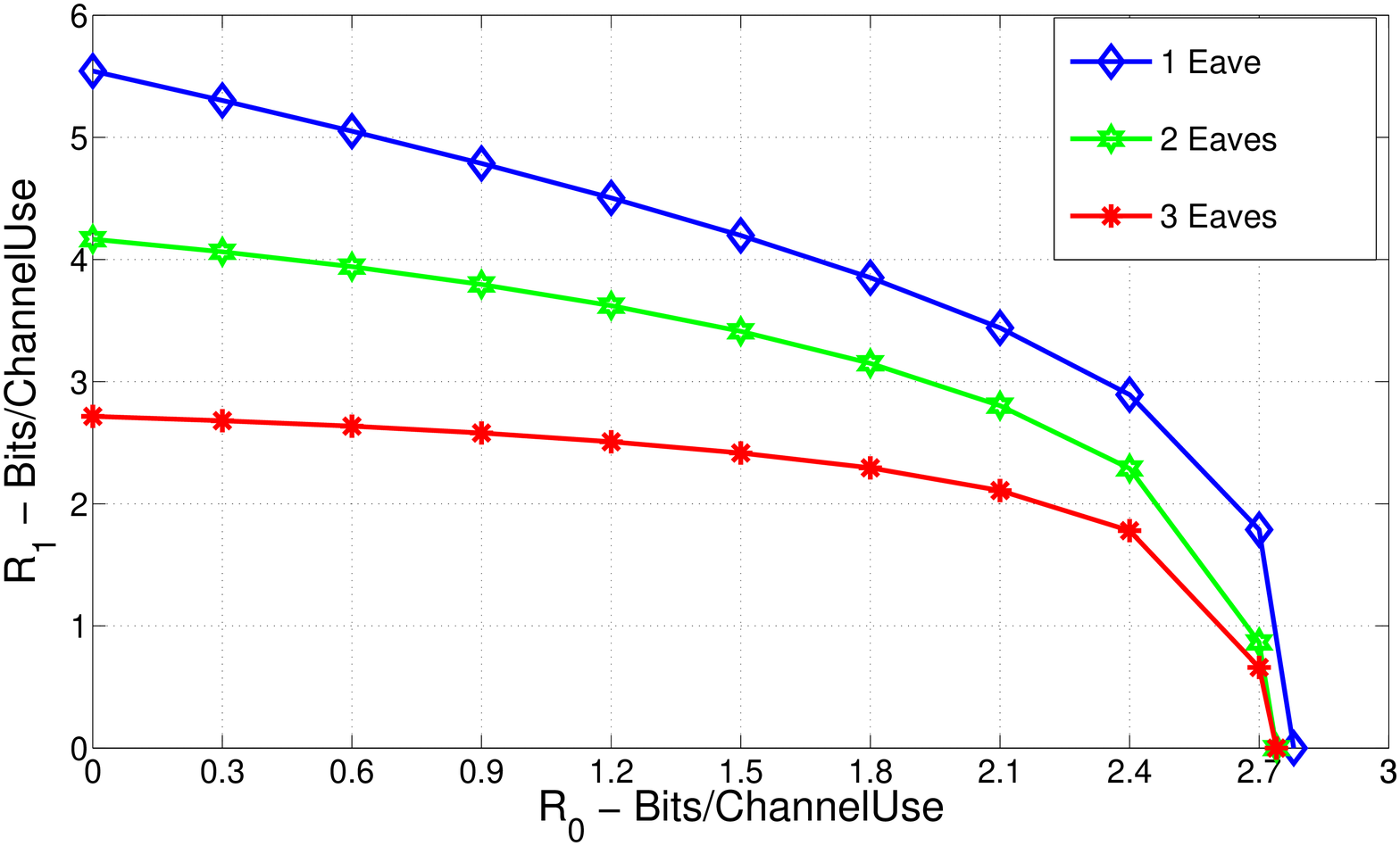,width=5.0cm, height=7.5cm}}
\centerline{\small{(b)}}\medskip
\end{minipage}
\caption{Secrecy rate in MISO broadcast channel for $N=2$, $K = 1$, $J=1,2,3$, and 
no artificial noise. (a) secrecy rate vs total power ($P_T$) with/without $W_{0}$, $R_0=1$; (b) {$R_1$ vs $R_0$} for 
$P_T=9$ dB.}
\label{fig2}
\end{figure}

Figure \ref{fig2}(a) shows the secrecy rate plots for MISO broadcast channel
as a function of total transmit power ($P_T$) when no artificial noise is injected.
The secrecy rates are plotted for the cases of 
with and without $W_{0}$. 
For the case with $W_{0}$, the information rate of $W_{0}$ is fixed at 
$R_0=1$. From Fig. \ref{fig2}(a), we observe that, for a given number of 
eavesdroppers, the secrecy rate degrades when $W_{0}$ is present.
Also, the secrecy rate degrades for 
increasing number of eavesdroppers. Figure \ref{fig2}(b)
shows the $R_{1}$ 
vs $R_0$ tradeoff, where $R_{1}$ is plotted as a function of $R_0$ for $K=1$,
$J=1,2,3$ at a fixed total power of $P_{T}=9$ dB and no artificial noise. 
It can be seen that as 
$R_0$ is increased, secrecy rate decreases. This is because the available 
transmit power for $W_{1}$ decreases as $R_0$ is increased. 
The point $2.7$ (approximately) on the $R_{0}$ axis where the secrecy rate drops to zero 
corresponds to $R^{max}_{0}$.

\begin{figure}[htb]
\begin{minipage}[b]{.48\linewidth}
\centering
\centerline{\epsfig{figure=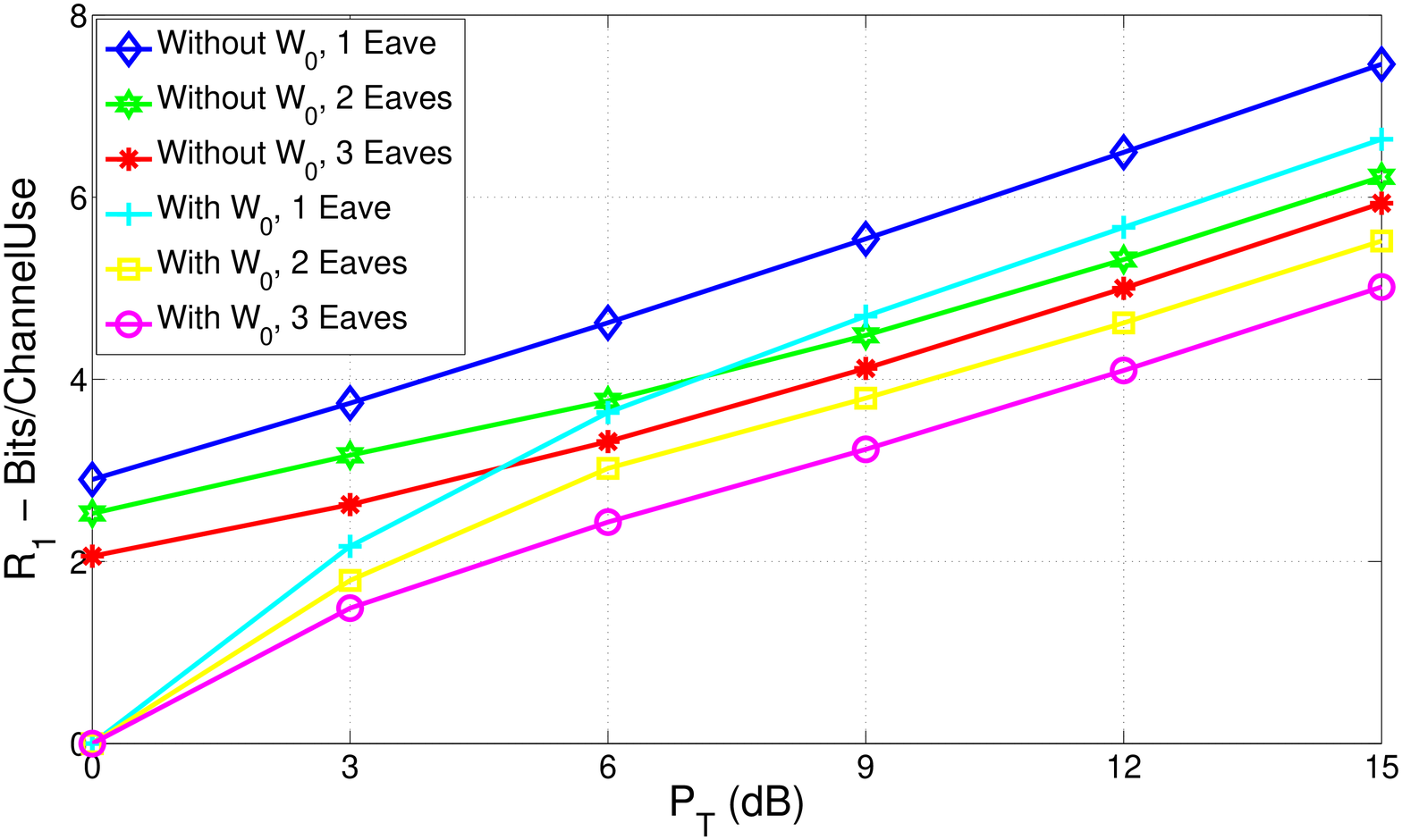,width=5.0cm,height=7.5cm}}
\centerline{\small{(a)}}\medskip
\end{minipage}
\hfill
\begin{minipage}[b]{0.48\linewidth}
\centering
\centerline{\epsfig{figure=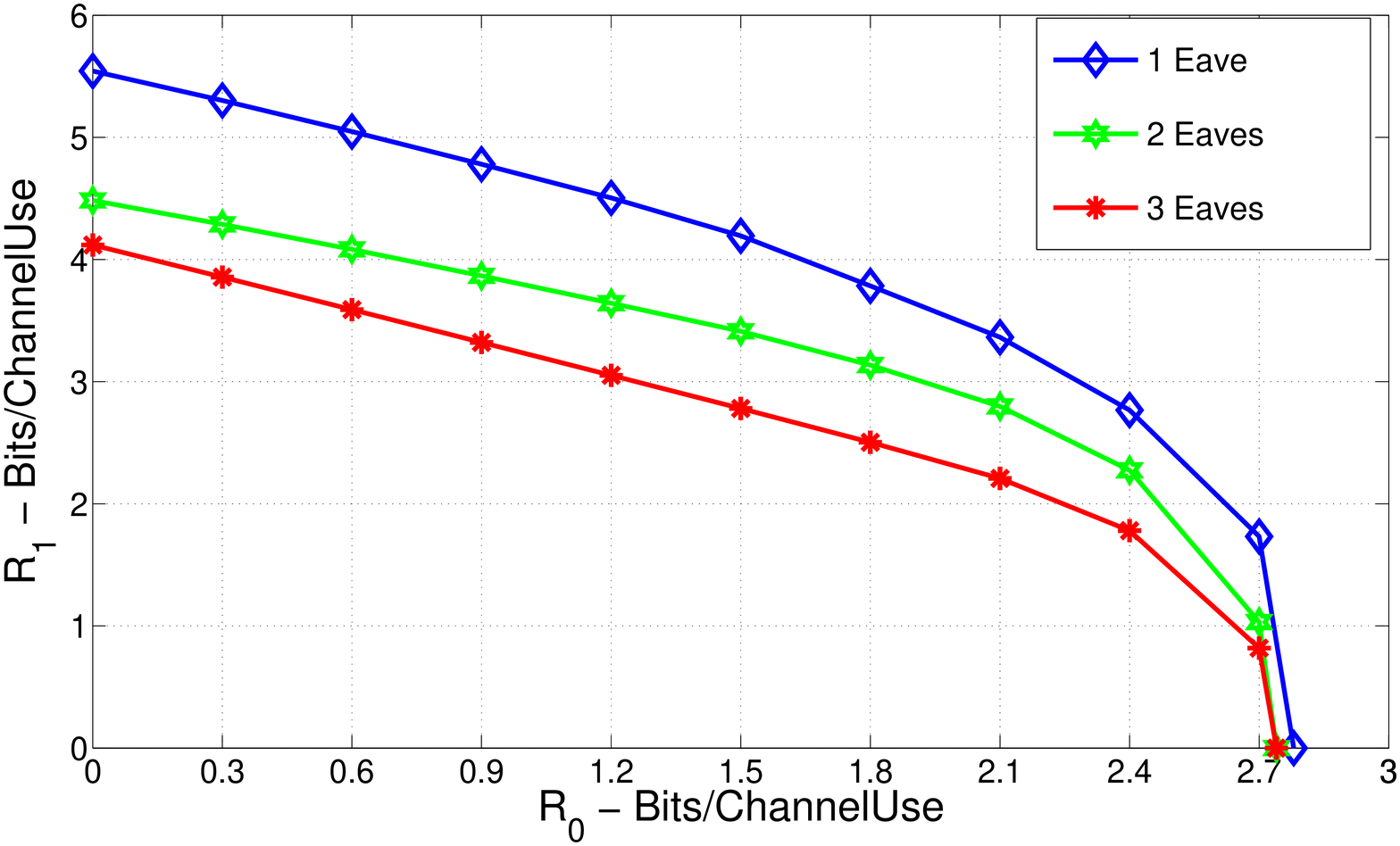,width=5.0cm, height=7.5cm}}
\centerline{\small{(b)}}\medskip
\end{minipage}
\caption{Secrecy rate in MISO broadcast channel for $N=2$, $K = 1$, $J=1,2,3$, and 
with artificial noise. (a) secrecy rate vs total power ($P_T$) with/without $W_{0}$, $R_0=1$; (b) {$R_1$ vs $R_0$} for 
$P_T=9$ dB.}
\label{fig3}
\end{figure}

Figure \ref{fig3}(a) 
shows the secrecy rate plots for MISO broadcast channel
as a function of total transmit power ($P_T$) when artificial noise is injected.
Similarly, Fig. \ref{fig3}(b)
shows the $R_{1}$ 
vs $R_0$ tradeoff with artificial noise, where $R_{1}$ is plotted as a function of $R_0$ for $K=1$,
$J=1,2,3$ at a fixed total power of $P_{T}=9$ dB.
We observe a significant improvement in secrecy rate as compared to Fig. \ref{fig2}(a) and Fig. \ref{fig2}(b)
when $J=2$ or $3$ eavesdroppers are present. When only one eavesdropper is present, artificial noise does not
help in improving the secrecy rate. This is due to the null signal beamforming by the source at 
the eavesdropper which is only possible when $J < N$.
Also, for the above channel conditions, we observe that
the solutions $\boldsymbol{Q}_{0}$ and $\boldsymbol{Q}_{1}$ 
obtained by solving the optimization problems (\ref{eqn44}) and (\ref{eqn62}) have rank 1.
\section{Conclusions}
\label{sec6}
We investigated transmitter optimization problem
in MISO broadcast channel with common and secret messages.
The source operates under a total power constraint. 
It also injects artificial noise
to improve the secrecy rate.
We obtained the optimum covariance matrices associated with
the common message, secret message, and artificial noise,
which maximized the achievable secrecy rate and simultaneously met the fixed rate $R_{0}$ for the common message.


\begin{thebibliography}{99}
\bibitem{ic0}
A. Wyner, ``The wire-tap channel,'' Bell. Syst Tech. J, vol. 54, no. 8,
pp. 1355-1387, Jan. 1975.

\vspace{0mm}
\bibitem{ic1}
I. Csiszar and J. Korner, ``Broadcast channels with confidential messages,'' 
{\em IEEE Trans. Inform. Theory}, vol. IT-24, pp. 339-348, May 1978.

\vspace{0mm}
\bibitem{ic2}
S. K. Leung-Yan-Cheong and M. E. Hellman, ``The Gaussian wire-tap channel,'' 
{\em IEEE Trans. Inform. Theory}, vol. IT-24, pp. 451-456, Jul. 1978.

\bibitem{ic3}
Y. Liang, H. V. Poor, and S. Shamai (Shitz), ``Information theoretic security,''
{\em Foundations and Trends in Communications and Information Theory}, NOW
Publishers, vol. 5, no. 4-5, 2009.

\vspace{0mm}
\bibitem{ic4}
F. Oggier and B. Hassibi, ``The secrecy capacity of the MIMO wiretap channel,'' 
{\em Proc. IEEE ISIT'2008}, July 2008.

\vspace{0mm}
\bibitem{ic11}
A. Khisti and G. Wornell, ``Secure transmission with multiple antennas-I: The
MISOME wiretap channel,'' {\em IEEE Trans. Inform. Theory}, vol. 56, no. 7,
pp. 3088-3104, Jul. 2010.

\vspace{0mm}
\bibitem{ic5}
A. Khisti and G. Wornell, ``Secure transmission with multiple antennas-II: The
MIMOME wiretap channel,'' {\em IEEE Trans. Inform. Theory}, vol. 56, no. 11,
pp. 5515-5532, Nov. 2010.

\vspace{0mm}
\bibitem{ic6}
 Y. Liang1, G. Kramer, H. V. Poor, and S. Shamai, ``Compound wiretap channels,'' 
{\em EURASIP Journ. on Wireless Commun. and Net.}, volume 2009, article ID 142374, 12 pages.
doi:10.1155/2009/142374
\vspace{0mm}
\bibitem{ic7}
H. D. Ly, T. Liu, and Y. Liang, ``Multiple-input multiple-output Gaussian 
broadcast channels with common and confidential messages,'' {\em IEEE Trans. 
Inform. Theory}, vol. 56, no. 11, pp. 5477-5487, Nov. 2010.

\vspace{0mm}
\bibitem{ic8}
E. Ekrem and S. Ulukus, ``Capacity region of Gaussian MIMO broadcast
channels with common and confidential messages,'' {\em IEEE Trans. Inform. 
Theory}, vol. 58, no. 9, pp. 5669-5680, Sep. 2012.

\vspace{0mm}
\bibitem{ic9}
R. Liu, T. Liu, and H. V. Poor, ``New results on multiple-input multiple-output
broadcast channels with confidential messages,'' {\em IEEE Trans. Inform. 
Theory}, vol. 59, no. 3, pp. 1346-1359, Mar. 2013.

\vspace{0mm}
\bibitem{ic20}
Q. Li and W. K. Ma, ``Spatially selective artificial-noise aided transmit
optimization for MISO multi-eves secrecy rate maximization,''
{\em IEEE Trans. Sig. Proc.}, vol. 61, no. 10, pp. 2704-2717, Mar 2013.

\vspace{0mm}
\bibitem{ic10}
S. Boyd and L. Vandenberghe, {\em Convex optimization}, Cambridge Univ. Press, 2004.

\end{thebibliography}
\end{document}